\documentclass[10pt,aps,prb,twocolumn,showpacs]{revtex4}

\usepackage{graphicx}

\begin{document}

\title{Spectral characteristics of noisy Josephson flux flow oscillator}
\author{E.A. Frolova}
\author{A.L. Pankratov}
\email{alp@ipm.sci-nnov.ru}
\author{M.Yu. Levichev}
\author{V.L. Vaks}
\affiliation{Institute for Physics of Microstructures of RAS, Nizhny Novgorod, 603950, Russia}

\begin{abstract}
The current-voltage and spectral characteristics of a flux flow oscillator (FFO), based on a long Josephson junction, are studied. The investigations are performed in the range of small bias currents and magnetic fields, where the FFO radiates a quasi-chaotic signal with extremely large radiation linewidth, and the displaced linear slope (DLS) is observed at the current-voltage characteristic. By direct numerical simulation of the sine-Gordon equation it is demonstrated that for large lengths of the Josephson junction or in the case of finite matching of the FFO with external waveguide system, the DLS with extremely large linewidth is transformed into Fiske steps with very narrow linewidth. While there is the common belief that the chaotic regime of the FFO is due to excitation of the internal oscillation modes in the "soft" fluxon chain, it is demonstrated that this regime is inspired by multiple reflections of the traveling waves from junction ends.
\end{abstract}

\maketitle

The flux flow oscillator\cite{ks} (FFO) based on a long Nb-AlOx-Nb Josephson junction of overlap geometry, generating a broadband quasi-chaotic signal, is a good candidate to be a noisy source for variety of applications, such as: nonstationary spectroscopy \cite{vks}, calibration of mixers, such as based on SIS (superconductor-insulator-superconductor) Josephson junctions, and  calibration of superconductive integrated receiver (SIR) \cite{sir},\cite{sir2}.
In the presence of an external magnetic field, the Josephson fluxons are formed and under the influence of a bias current start to move from one end of the junction to another one and are radiated at the border of the junction. Depending on the value of a magnetic field and a bias current supplied to the FFO, three various modes of generation can be outlined. At small magnetic fields, of order of a critical field at which fluxons start to enter the junction, the chaotic generation with very large width of a spectral line, which can reach several GHz at frequency of generation from 50 to 200 GHz, is observed \cite{ukh}. Further, with increase in a magnetic field and frequency of generation (from 200 to approximately 400 GHz), almost vertical Fiske steps are seen at the current-voltage characteristic (IVC)\cite{nei}-\cite{jaw}. Here, due to resonant mode, the spectral linewidth is smaller than 1 MHz, but thus, because of the Fiske steps, there is no possibility of smooth frequency tuning that complicates application of this mode for spectroscopic measurements. With further increase of the magnetic field and the oscillation frequency in the range from 450 to 700 GHz, the continuous flux flow steps  are observed at IVC\cite{nei}-\cite{jaw}. Here the frequency can continuously be tuned in the whole range leading to the width of a spectral line from 1 to 50 MHz.

It is widely accepted in the literature that in the range of FFO frequency from 50 to 200 GHz the so-called displaced linear slope (DLS) is observed at current-voltage characteristic \cite{ukh},\cite{SP},\cite{cik},\cite{mart}, and the generation linewidth in this range is extremely broad. It is usually attributed to excitation of the internal oscillation modes in the "soft" fluxon chain at weak magnetic fields \cite{ukh},\cite{SP}. However, most of experimental and especially theoretical works dealt with the case where the matching between FFO and external wave-guide system was pure or even absent \cite{ukh}-\cite{MJ}.

The aim of the present paper is the investigation of an influence of thermal fluctuations, a finite value of RC load and a length of the FFO on the spectral line in this quasi-chaotic regime of FFO radiation. We demonstrate that for sufficiently long junctions or in the case of good matching, which can be modeled by RC-load at FFO ends, the DLS with broad linewidth is transformed into Fiske steps with extremely narrow linewidth. Therefore, the appearance of regime of chaotic oscillations can be explained by multiple reflections of the traveling waves from junction ends.

It is known that all basic properties of the FFO can be described in the frame of the sine-Gordon equation:
\begin{equation}
{\phi}_{tt}+\alpha{\phi}_{t}
-{\phi}_{xx}=\beta{\phi}_{xxt}+\eta-\sin (\phi)+\eta_f(x,t)
\end{equation}
where indices $t$ and $x$ denote temporal and spatial derivatives, $\varphi$ is the phase order parameter. Space and time are normalized to the Josephson penetration length
$\lambda _{J}$ and to the inverse plasma frequency
$\omega_{p}^{-1}$, respectively, $\alpha={\omega_{p}}/{\omega_{c}}$
is the damping parameter, $\omega_p=\sqrt{2eI_c/\hbar C}$,
$\omega_{c}=2eI_cR_{N}/\hbar$, $I_c$ is the critical current, $C$ is
the JTJ capacitance, $R_N$ is the normal state resistance, $\beta$
is the surface loss parameter, $\eta$ is the dc overlap bias
current density, normalized to the critical current density $J_c$,
and $\eta_f(x,t)$ is the fluctuational current density. If the
critical current density is fixed and the fluctuations are treated
as white Gaussian noise with zero mean, its correlation function is:
$\left<\eta_f(x,t)\eta_f(x',t')\right>=2\alpha\gamma \delta
(x-x^{\prime})\delta (t-t^{\prime})$, where $\gamma = I_{T} /
(J_{c}\lambda_J)$ is the dimensionless noise intensity,
$I_{T}=2ekT/\hbar$ is the thermal current, $e$ is the electron
charge, $\hbar$ is the Planck constant, $k$ is the Boltzmann
constant and $T$ is the temperature.

The boundary conditions, that simulate simple RC-loads, see Ref.s
\cite{Parment,pskm,p,pa}, have the form:
\begin{eqnarray}\label{x=0}
\phi(0,t)_{x}+r_L c_L\phi(0,t)_{xt}-c_L\phi(0,t)_{t t}+\\
\beta r_L c_L\phi(0,t)_{xtt}+\beta\phi(0,t)_{x
t}=\Gamma, \nonumber \\ \phi(L,t)_{x}+r_R
c_R\phi(L,t)_{x t}+c_R\phi(L,t)_{t t}+ \label{x=L} \\  \beta r_R
c_R\phi(L,t)_{xtt}+\beta\phi(L,t)_{x t}=\Gamma.
\nonumber
\end{eqnarray}
Here $\Gamma$ is the normalized magnetic field, and $L$ is the
dimensionless length of the FFO in units of $\lambda _{J}$. The dimensionless capacitances and
resistances, $c_{L,R}$ and $r_{L,R}$, are the FFO RC-load placed at
the left (output) and at the right (input) ends, respectively.

\begin{figure}[ht]
\resizebox{1\columnwidth}{!}{
\includegraphics{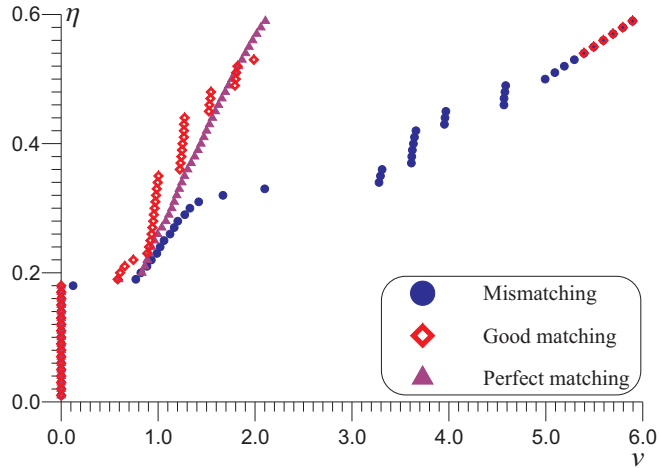}}
{\caption{Current-voltage characteristic of FFO for various matching at junction ends for $\alpha=0.1$, $L=10$, $\Gamma=1.7$, $\gamma=0.001$. Circles -- mismatching, $r_{L}=r_{R}=100$; diamonds -- good matching at the radiating end, $r_{L}=2$, $r_{R}=100$; triangles -- perfect matching, $r_{L}=r_{R}=1$.}
\label{CV}}
\end{figure}

\begin{figure}[ht]
\resizebox{1\columnwidth}{!}{
\includegraphics{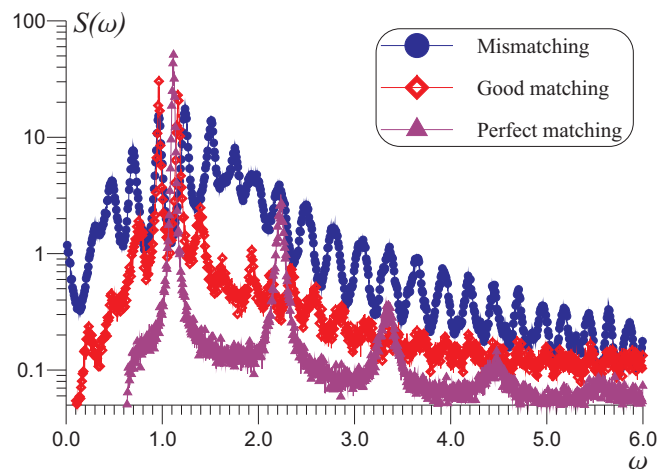}}
{\caption{Spectral density of FFO for various matching of junction ends for $\alpha=0.1$, $L=10$, $\Gamma=1.7$, $\gamma=0.1$, $\eta=0.3$. Circles -- mismatching, $r_{L}=r_{R}=100$; diamonds -- good matching at the radiating end, $r_{L}=2$, $r_{R}=100$; triangles -- perfect matching, $r_{L}=r_{R}=1$.}
\label{FigSP1}}
\end{figure}
The computer simulations are performed at the same parameters, as in work\cite{ukh}:  $\alpha$=0.1, $\beta$=0, the external magnetic field $\Gamma$=1.5-1.9, $\beta=0$, the junction length $L=10\div 80$. At these parameters, and for absence of the matching with external waveguide system, e.g., $r_{L}=r_{R}=100$, $c_{L}=c_{R}=10$, for $L=10$, the specific displaced linear slope (DLS) appears at I-V characteristics (Fig. \ref{CV}, circles) where chaotic generation is observed (Fig. \ref{FigSP1}, circles). It should be noted that, in difference with Fig. 5 of Ref. \cite{ukh} we do not observe high-frequency noisy oscillations of spectra: we perform ensemble averaging of spectra over 500 realizations and the only envelope curve remains, which looks very similar to the one in Ref. \cite{ukh}. All curves in Fig. \ref{FigSP1} are computed for $\eta=0.3$, see Fig. \ref{CV}. The account of small noise in the system does not change the character of the chaotic generation: the IVC remains nearly the same in the range $v=0.8\div1.4$, Fig. \ref{CV}, and the spectrum remains broad enough. However, matching the FFO with external waveguide system makes an essential impact on the IVC and the spectral density of a generated signal. At good enough matching at the radiating end, $r_{L}=2$, $r_{R}=100$, the series of Fiske steps appear at the IVC (Fig. \ref{CV}, diamonds), the width of the spectral line decreases essentially (Fig. \ref{FigSP1}, diamonds). With further matching improvement, $r_{L}=r_{R}=1$ -- perfect matching, the corresponding IVC merges into the continuous curve (which qualitatively looks very similar to the DLS in the mismatching case, only the working frequency range is larger), see Fig. \ref{CV}, triangles, and the generation becomes quasi-monochromatic (Fig. \ref{FigSP1}, triangles). In real devices, such smoothing of the IVCs can also be due to surface losses \cite{pskm,MJ} and self-pumping effect \cite{SP,pskm,MJ}, which, however, become significant for larger frequencies only \cite{pskm}. In both cases of good and perfect matching the width of the spectral line is defined by the intensity of thermal noise of the Josephson junction \cite{p,pa}. Thus, the quasi-chaotic generation predicted and observed in work\cite{ukh} arises due to numerous reflections of Josephson vortices from junction ends. In a real situation where the FFO is matched with an external line for effective transfer of radiation, the chaotic mode of generation is replaced by the quasi-monochromatic one.

\begin{figure}[ht]
\resizebox{1\columnwidth}{!}{
\includegraphics{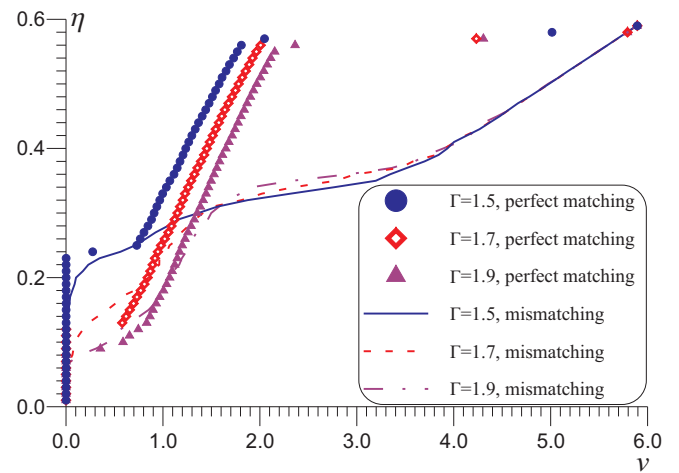}}
{\caption{I-V characteristics of the FFO at various values of external magnetic field, and various conditions of the matching; $L=10$, $\gamma=0.1$.}
\label{FigCV1}}
\end{figure}

\begin{figure}[ht]
\resizebox{1\columnwidth}{!}{
\includegraphics{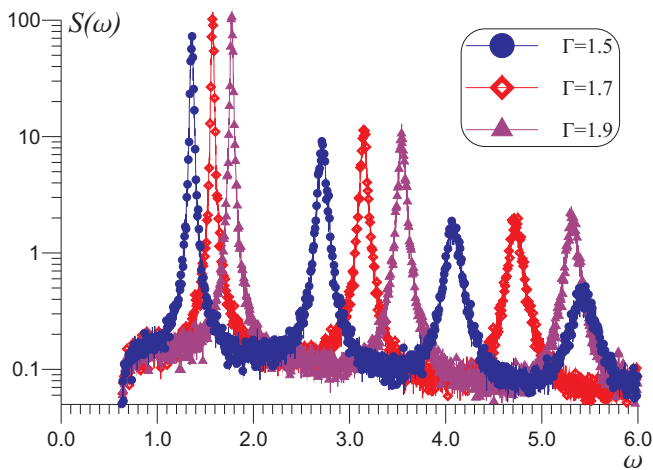}}
{\caption{Power spectral density of the FFO at various external magnetic fields $\Gamma=1.5$, $\Gamma=1.7$ and $\Gamma=1.9$; $L=10$, $\gamma=0.1$, $\eta=0.45$.}
\label{FigSP2}}
\end{figure}

The matching at the junction ends essentially influences the working frequency range.
This is especially important for smaller magnetic fields, where in the mismatched case the generation step is nearly absent, see Fig. \ref{FigCV1} for $\Gamma=1.5$. We note that the DLS $v=0.8\div1.4$ is very stable to the effect of fluctuations, while the steps in the range $v=3\div 5$ are completely smoothed for $\gamma=0.1$, compare with Fig. \ref{CV} for $\gamma=0.001$, the DLS in Fig. \ref{FigCV1} remains nearly unaffected by noise.
Due to matching, for small magnetic fields $\Gamma=1.5\div1.9$ the spectra become very narrow (Fig. \ref{FigSP2}, circles, diamonds and triangles for fields $\Gamma$=1.5$, \Gamma$=1.7 and $\Gamma$=1.9, respectively). The frequency spacing between the peaks of the spectral density is proportional to the voltage at the point of the current-voltage characteristic of the junction. Thus, independently of the choice of the external magnetic field value, for the well-matched case the chaotic generation disappears. Therefore, it can be recommended for practical applications to place Chebyshev transformers not only at the output end \cite{ks},\cite{SP}, but also at the input end of the FFO to improve matching and to consequently suppress reflections of traveling waves from junction ends. This can lead to significant smoothing of IV curves and to continuous frequency tuning in all working range from 50 to 750 GHz.

\begin{figure}[ht]
\resizebox{1\columnwidth}{!}{
\includegraphics{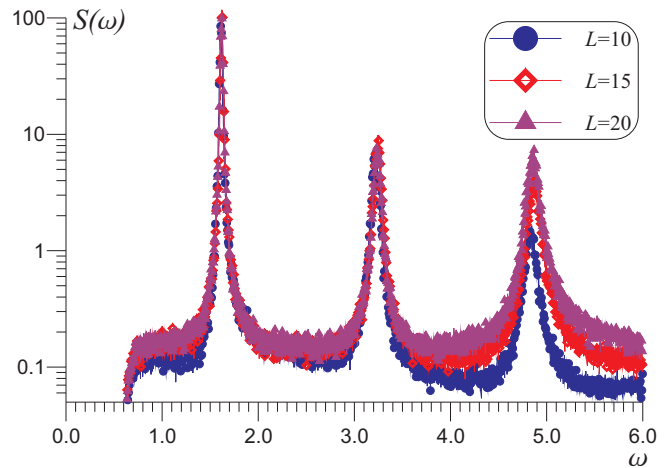}}
{\caption{The spectral density of the FFO radiation for the perfectly-matched case at both junction ends at various lengths of the junction  $L=10$, $L=15$ and $L=20$; $\gamma=0.1$, $\eta=0.4$.}
\label{FigSP4}}
\end{figure}

Comparison of power spectral densities for various lengths of the junction $L=10$, $L=15$ and $L=20$, in the perfectly-matched case (Fig. \ref{FigSP4}, circles, diamonds and triangles, respectively) shows qualitative independence from the length of the junction.

\begin{figure}[ht]
\resizebox{1\columnwidth}{!}{
\includegraphics{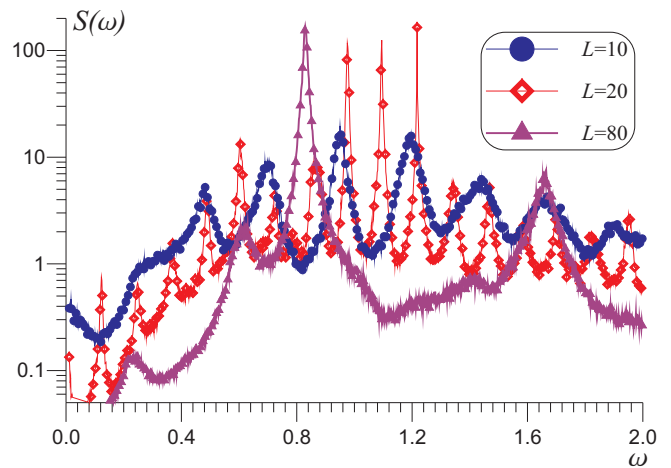}}
{\caption{The spectral density in the mismatching case for different lengths of the junction $L=10$, $L=20$ and $L=80$; $\gamma=0.01$, $\eta=0.25$.}
\label{FigSP5}}
\end{figure}

\begin{figure}[ht]
\resizebox{1\columnwidth}{!}{
\includegraphics{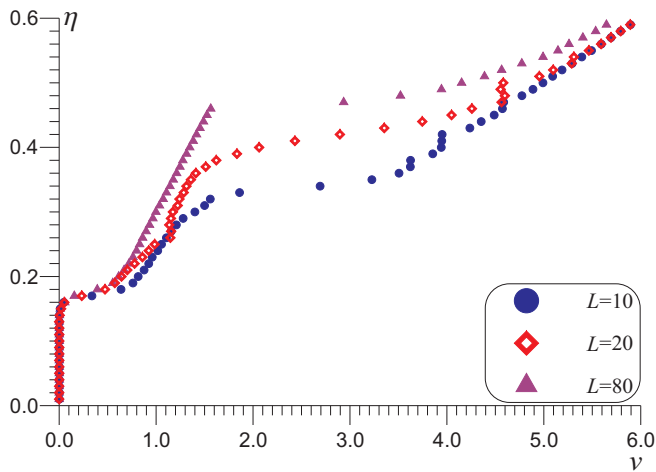}}
{\caption{The I-V characteristic at various lengths of the junction $L=10$, $L=20$, $L=80$, mismatching case; $\gamma=0.01$.}
\label{FigCVL}}
\end{figure}
The situation is completely different for the unloaded case. As one can see from Fig. \ref{FigSP5}, $\Gamma=1.7$, with increase of $L$ the chaotic generation for $L=10$ transforms to the multi-harmonic generation for $L=20$, and to the harmonic generation with one strongly dominating peak for $L=80$. Correspondingly, the increase of the FFO length also significantly affects the current-voltage characteristic, increasing the IVC slope, see Fig. \ref{FigCVL}.

Thus, the use of the chaotic mode of generation of the FFO with a broad spectrum has a number of restrictions that does not give the opportunity of its use for creation of subTHz non-stationary spectrometers. However, other modes of the FFO can be used, where quasi-monochromatic generation with broad linewidth is also observed (for example, in more important range of frequencies from 450 to 700 GHz), besides for additional broadening of the spectrum an additional stochastization of bias current and magnetic field of the FFO can be used.

In the present paper it is shown that the power spectral density of the FFO essentially depends on the junction length and conditions of the RC-load matching at the ends of the Josephson junction. The special case of extremely broadband chaotic generation of the Josephson flux flow oscillator is observed only at the mismatching of the FFO with external waveguide system and small junction lengths. Therefore, the appearance of regime of chaotic oscillations can be explained by multiple reflections of the traveling waves from junction ends rather than simply by excitation of the internal oscillation modes in the "soft" fluxon chain at weak magnetic fields \cite{ukh},\cite{SP}.

The work is supported by the RFBR (projects 09-02-00491 and 09-02-97085).

\end{document}